# SrRuO$_3$ under tensile strain: Thickness-dependent electronic and magnetic properties

*Yuki K. Wakabayashi,[1,a] Masaki Kobayashi,[2,3] Yuichi Seki,[3] Kohei Yamagami,[4] Takahito Takeda,[3] Takuo Ohkochi,[4,5] Yoshitaka Taniyasu,[1] Yoshiharu Krockenberger,[1] and Hideki Yamamoto[1]*

[1]NTT Basic Research Laboratories, NTT Corporation, Atsugi, Kanagawa 243-0198, Japan
[2]Center for Spintronics Research Network, The University of Tokyo, 7-3-1 Hongo, Bunkyo-ku, Tokyo 113-8656, Japan
[3]Department of Electrical Engineering and Information Systems, The University of Tokyo, Bunkyo, Tokyo 113-8656, Japan
[4]Japan Synchrotron Radiation Research Institute (JASRI), 1-1-1 Kouto, Sayo, Hyogo, 679-5198, Japan
[5]Laboratory of Advanced Science and Technology for Industry, University of Hyogo, Kamigori, Hyogo 678-1205, Japan

a)Author to whom correspondence should be addressed: yuuki.wakabayashi@ntt.com

**ABSTRACT**
The burgeoning fields of spintronics and topological electronics require materials possessing a unique combination of properties: ferromagnetism, metallicity, and chemical stability. SrRuO$_3$ (SRO) stands out as a compelling candidate due to its exceptional combination of these attributes. However, understanding its behavior under tensile strain, especially its thickness-dependent changes, remains elusive. This study employs machine-learning-assisted molecular beam epitaxy to investigate SRO films with thicknesses from 1 to 10 nm. This work complements the existing focus on compressive-strained SRO, opening a new avenue for exploring its hitherto concealed potential. Using soft X-ray magnetic circular dichroism, we uncover an intriguing interplay between film thickness, electronic structure, and magnetic properties. Our key findings reveal an intensified localization of Ru 4$d$ $t_{2g}$-O 2$p$ hybridized states at lower thicknesses, attributed to the weakened orbital hybridization. Furthermore, we find a progressive reduction of magnetic moments for both Ru and O ions as film thickness decreases. Notably, a non-ferromagnetic insulating state emerges at a critical thickness of 1 nm, marking a pivotal transition from the metallic ferromagnetic phase. These insights emphasize the importance of considering thickness-dependent properties when tailoring SRO for next-generation spintronic and topological electronic devices.

## I. INTRODUCTION

Due to its unique combination of ferromagnetism, metallicity, chemical stability, and compatibility with other perovskite-structured oxides, the 4$d$ itinerant ferromagnetic perovskite SrRuO$_3$ (SRO), whose bulk Curie temperature ($T_C$) is 165 K, has been extensively studied.[1-9] The recent observations of Weyl fermions in it has sparked renewed attention for investigating unique quantum transport phenomena and possible applications in the field of topological electronics.[10-14] In addition, a ferromagnetic state at a single monolayer SRO has been reported,[15] making it a unique candidate for realizing single-layer magnetic Weyl semimetals. Additionally, it has been reported that SRO exhibits spin-related transport phenomena such as tunneling magnetoresistance[16] and current-injected magnetization switching,[17] making SRO promising for applications in spintronics.

   The successful integration of SRO into spintronics and topological electronics necessitates comprehensive consideration of its strain-dependent electrical and magnetic properties at heterointerfaces, particularly under the ubiquitous presence of epitaxial strain arising from lattice mismatch. While the thickness dependence of these properties under compressive strain on SrTiO$_3$ (STO) (001) substrates has been extensively investigated,[15,18-22] a critical knowledge gap exists concerning SRO under tensile strain. Notably, in compressive-strained SRO films, a well-defined critical thickness of ~ 2 nm marks the transition to a non-ferromagnetic, insulating state. However, the strong coupling between lattice, electron, and spin degrees of freedom in SRO[23-27] casts doubt on the direct extrapolation of this behavior to tensile strain scenarios. Therefore, a systematic investigation of the thickness-dependent evolution of electronic and magnetic properties in tensile-strained SRO films is an urgent and paramount endeavor.

   In this study, we investigated the magnetic and electrical properties of tensile-strained SRO films with thicknesses $t$ = 1–10 nm grown by machine-learning-assisted molecular beam epitaxy (ML-MBE).[28-30] The transport properties were characterized by magnetotransport measurements. The element-specific electronic structure and magnetic properties of the Ru 4$d$ and O 2$p$ states were clarified using soft X-ray absorption spectroscopy (XAS) and X-ray magnetic circular dichroism (XMCD).[31-34] Our investigation reveals a pronounced localization of the Ru 4$d$ $t_{2g}$-O 2$p$ hybridized states with decreasing film thickness ($t$). The magnetic moments of Ru and O ions decreased with decreasing $t$ and the SRO film became nonferromagnetic at $t$ = 1 nm. The increased localization also resulted in insulating electrical properties of the SRO films with $t$ = 1 and 2 nm. These results indicate that the localization due to disorders near the interface[35,36] weakens orbital hybridization and causes the change from a ferromagnetic metallic state in thick films to a nonferromagnetic insulating state at $t$ = 1 nm.

## II. METHODS

We grew epitaxial SRO films having 0.8% tensile strain with $t$ of 1, 2, and 10 nm on TSO (110) substrates [Fig. 1(a)] in an MBE system equipped with multiple e-beam evaporators for Sr and Ru. The oxidation during growth was carried out with a mixture of ozone (O$_3$) and O$_2$ gas (~15% O$_3$ + ~85% O$_2$). All SRO films were prepared under the same growth conditions as in Ref. 27, in which detailed crystallographic analyses by $\theta$−2$\theta$ X-ray diffraction and high-resolution X-ray reciprocal space mapping for the SRO film with $t$ = 60 nm on TSO (110) substrate were reported. The growth parameters were optimized by Bayesian optimization, a machine learning technique.[28,37] The sharp streaky reflection high-energy electron diffraction patterns confirmed in-plane epitaxial growth of the SRO films. Further information about the MBE setup and preparation of the substrates is described elsewhere.[28,38]

Magnetotransport was measured using a standard four-probe method with Ag electrodes deposited on the SRO surface in a DynaCool physical property measurement system (PPMS).

The XAS and XMCD measurements were performed at the helical undulator beamline BL25SU of SPring-8.[39,40] The monochromator resolution $E/\Delta E$ was over 5,000 at the Ru $M_{2,3}$ edges. The designed beam spot size was 10 × 100–200 μm$^2$. For the XMCD measurements, absorption spectra for circularly polarized X-rays with the photon helicity parallel ($\mu^+$) or antiparallel ($\mu^-$) to the spin polarization were obtained by reversing the photon helicity at each photon energy $h\nu$ and recording them in the total-electron-yield (TEY) mode. The measurement temperature was 30 K. All XAS and XMCD spectroscopy measurements were performed at 1.92 T, the maximum applied magnetic field in the system.[39,40] The $\mu^+$ and $\mu^-$ spectra at the Ru $M_{2,3}$ edges and O $K$ edge were taken under both positive and negative applied magnetic fields and averaged to eliminate spurious dichroic signals. To estimate the integrated values of the XAS spectra at the Ru $M_{2,3}$ edges, hyperbolic tangent functions as background were subtracted from the spectra. To measure the perpendicular and in-plane magnetic moments, the angles of the external magnetic fields and incident X-rays measured from the sample surfaces were set to $\theta = 90°$ and 20°, respectively [Figs. 1(b) and 1(c)].

**RESULTS**

Figure 2(a) shows the temperature dependence of the longitudinal resistivity $\rho$ for 0.8% tensile-strained epitaxial single-crystalline SRO films with thicknesses $t$ of 1, 2, 4, and 10 nm on TbScO$_3$ (TSO) (110) substrates grown by ML-MBE. TSO has the GdFeO$_3$ structure, which is a distorted perovskite structure with the (110) face corresponding to the pseudocubic (001) face. The $\rho$ values of the films with $t = 10$ and 4 nm decreases with decreasing temperature $T$, indicating that these films are metallic in the whole measurement temperature range. Similarly, the $\rho$ value of the 2-nm film decreases upon decreasing temperature from 300 K down to 26.5 K, but for $T < 26.5$ K, $\rho$ increases with decreasing temperature, indicating insulating behavior due to weak localization of charge carriers in the low-temperature region.[15,22] With a further decrease in $t$ to 1 nm [inset in Figure 2(a)], the film shows insulating behavior (d$\rho$/d$T$ is negative) for the whole measurement temperature range. In addition, the thinner the film is, the larger the $\rho$ value becomes at each temperature. These results suggest that the interface-driven disorder causes increased resistivity and carrier localization. The residual resistivity ratio (RRR) [= $\rho(300$ K$)/\rho(2$ K$)$] of the SRO film with $t = 10$ nm was 10.1. This value is smaller than that for the -0.7% compressive-strained SRO film on STO (001) substrate with $t = 10$ nm (RRR = 21) grown by ML-MBE reported in Ref. 22. Furthermore, the -0.7% compressive-strained SRO/STO film with $t = 2$ nm does not show insulating behavior at low temperatures.[22] While the magnitude of the absolute strain experienced by compressive-strained SRO/STO and tensile-strained SRO/TSO films may be comparable, the former exhibits a significantly higher RRR. This discrepancy, which exists even though the growth conditions are optimized by Bayesian optimization in both strain cases, highlights the profound influence of strain type on SRO's electrical properties. Notably, the emergence of an insulating state at 2 nm only in tensile-strained SRO further underscores this distinction. Considering the importance of minimizing Joule heating and energy loss in heteroepitaxial devices, these findings suggest that compressive-strained SRO, with its higher conductivity and metallic behavior at thinner thicknesses, presents a more suitable choice for metallic ferromagnetic oxide electrodes. When $t = 10$ or 4 nm, the $\rho$ vs $T$ curves show a clear kink [Fig. 2(a)], at which the ferromagnetic transition occurs and spin-dependent scattering is suppressed.[4] To highlight this transition, Curie temperature $T_C$ at which the d$\rho$/d$T$ curves show a peak is also shown in black arrows in Fig.

2(a). Note that, generally, $T_C$ determined from $d\rho/dT$ is a few kelvins lower than the values measured from the temperature dependence of the magnetization.[41] The $T_C$ value decreases with decreasing $t$ from 10 nm to 2 nm. This criterion to define $T_C$ cannot be applied to the 1-nm film due to the insulating behavior. The lower $T_C$ values in the thinner films can be attributed to the degradation of the exchange interaction in SRO. The degradation should come from disorders near the interface.

We also performed magnetotransport measurements. Figure 2(b) shows the $t$ dependence of the magnetoresistance (MR) $[(\rho(\mu_0H) - \rho(0\,\text{T}))/\rho(0\,\text{T})]$, where the magnetic field $\mu_0H$ was applied in the out-of-plane [110] direction of the TSO substrate at 30 K. The SRO film with $t = 1$ nm could not be measured since $\rho$ at 30 K is larger than the measurement range of the DynaCool PPMS. For the SRO films with $t = 10$, 4, and 2 nm, negative anisotropic magnetoresistance (AMR),[42] which is proportional to the relative angle between the electric current and the magnetization, is clearly observed. This is in contrast to the positive AMR observed in compressive-strained SRO/STO films.[4,22] According to the extended Campbell-Fert model, the negative AMR is not observed for majority spin conduction in ferromagnets; thus, the negative AMR observed here is thought to come from the minority spin conduction.[43] The AMR peak position, which corresponds to the coercive field $H_c$ of SRO,[11] becomes larger with decreasing $t$. Since the magnetic domains tend to be pinned by grain boundaries and other defects, the large $H_c$ in thinner films also suggests poor crystalline quality near the interface. This result also supports the scenario that the disorders near the interface degrade the exchange interaction in SRO.

To get detailed insights into the thickness-dependent electronic structures and magnetic properties, we carried out XAS and XMCD measurements, which are tools sensitive to the element-specific electronic structure and magnetic properties.[31-34] First, we show the thickness-dependent XAS and XMCD results taken in the in-plane configuration ($\theta = 20°$) [Fig. 1(c)]. Figures 3(a)–3(c) shows XAS and XMCD spectra at the Ru $M_{2,3}$ edge with $t = 10$, 2, and 1 nm, taken at $\theta = 20°$. For all the films, the Ru $M_3$ and $M_2$ XAS absorptions from the Ru $3p_{3/2}$ and $3p_{1/2}$ core levels into the Ru $4d$ states are observed at around 463 and 485 eV.[44-46] On the other hand, the Ru $M_3$ and $M_2$ XMCD absorption peaks are observed only at $t = 10$ and 2 nm, indicating that the SRO films with $t = 10$ and 2 nm are ferromagnetic at 30 K, while the SRO film with 1 nm is nonferromagnetic. Together with the magnetotransport, these results show a change from ferromagnetic metal to nonferromagnetic insulator between $t = 2$ and 1 nm. As shown in Figs. 3(d)–3(f), the O $K$-edge XAS and XMCD spectra represent the unoccupied electronic structures hybridized with the O $2p$ states and the O $2p$ orbital magnetic moments [Figs. 3(d)–3(f)]. The absorption peak at 528.9 eV arises from the transition to the coherent part of the Ru $4d$ $t_{2g}$ states.[47] The spectra in the energy range of 529.5–531.5 eV correspond to the incoherent part of the Ru $4d$ $t_{2g}$ states, as reported by Takizawa et al[47] and Okamoto et al.[44] As with the Ru $M_{2,3}$ edge, the O $K$-edge XMCD peaks, which represent the substantial orbital magnetic moments of the O $2p$ states, were observed only at ferromagnetic states at $t = 10$ and 2 nm and not at 1 nm. The Ru $4d$ $t_{2g}$ peak in the O $K$-edge XMCD spectrum reveals a significant O $2p$ orbital magnetic moment due to a charge transfer from the O $2p$ states to the Ru $4d$ states. This suggests the formation of Ru $4d$-O $2p$ spin-polarized bands at the Fermi level ($E_F$).[11,46]

The large intensity of the coherent Ru $4d$ $t_{2g}$ peak at $t = 10$ nm indicates the high itineracy and long lifetime of the hybridized Ru $4d$ $t_{2g}$-O $2p$ states.[46,48] To investigate the variation of the itineracy of the hybridized state, we plotted the Ru $4d$ $t_{2g}$ peak intensity in the O $K$-edge XAS spectrum normalized at 531 eV, which corresponds the absorption energy for the incoherent part of the Ru 4$d$ $t_{2g}$ states, as a function of $t$ (blue triangles in Fig. 4). In Fig. 4, the data reported in Ref. 49 are also plotted for $t = 60$ nm. The coherent peak exhibits reduced intensity as $t$ decreases, suggesting a decrease in the lifetimes of quasiparticles within the hybridized Ru 4$d$ $t_{2g}$-O 2$p$ states. This decrease in intensity also indicates a strengthening of the localization of the hybridized states, attributed to disorders near the interface.[35,36] The increased localization with decreasing $t$ is thought to be the origin of the insulating electrical properties observed in Fig. 2(a). An increase in localization should result in lower hybridization strength between the Ru 4$d$ $t_{2g}$ and O 2$p$ states by shrinking the wave function. We also clarified the change in the magnetic moment accompanied by the weakening of the hybridization by evaluating the total magnetic moment ($M = m_{orb} + m_{spin}$) of Ru in the in-plane configuration ($\theta = 20°$), $M_\parallel^{Ru}$, and the $m_{orb}$ of O in the in-plane configuration ($\theta = 20°$), $m_{orb,\parallel}^{O}$. According to the XMCD sum rules,[31] the integrated intensity of XMCD divided by that of XAS (XMCD/XAS integrated intensity) at the Ru $M_2$ edge is proportional to the total magnetic moment $M$ of Ru.[50] Therefore, the $M$ of Ru was evaluated by integrating the Ru $M_2$-edge XMCD/XAS intensity from 471 to 495 eV. Similarly, the $m_{orb}$ of O was evaluated by integrating the O $K$-edge XMCD/XAS intensity[50,51] from 527 to 531 eV, where finite O $K$-edge XMCD intensity appears [Figs 3(d) and 3(e)]. As shown in Fig. 4, the $M_\parallel^{Ru}$ (green circles) and $m_{orb,\parallel}^{O}$ (red circles) decrease with the decrease in the O $K$-edge Ru 4$d$ $t_{2g}$ peak intensity (blue triangles) as $t$ decreases from 60 to 2 nm. In Fig. 4, the $M_\parallel^{Ru}$ and $m_{orb,\parallel}^{O}$ values are normalized by those at $t = 60$ nm to see the thickness dependence. This result means that the orbital magnetic moment of oxygen decreases as a result of reduced charge transfer through orbital hybridization and that the Ru 4$d$ $t_{2g}$-O 2$p$ hybridization is important for the ferromagnetic ordering of Ru. When $t$ further decreases to 2 nm, the O $K$-edge Ru 4$d$ $t_{2g}$ peak intensity decreases to 0.27 times that of 60 nm, and the XMCD peak disappears at $t = 1$ nm.

To examine how the magnetic anisotropy changes with $t$, we also performed the XMCD measurements in the perpendicular ($\theta = 90°$) configuration. Figures 5(a) and 5(b) show the normalized XMCD spectra at the Ru $M_{2,3}$ and O $K$ edges with $t = 10$ and 2 nm at $\theta = 90°$ and $\theta = 20°$. The $m_{orb}/m_{spin}$ ratio of Ru is exclusively derived from the ratio of the XMCD absorption intensities at the Ru $M_3$ edge to $M_2$ edge, as dictated by the XMCD sum rules. The substantial overlap in the Ru $M_{2,3}$-edge spectra means that the Ru 4$d$ states have a constant $m_{orb}/m_{spin}$ of 0.081, as determined by the XMCD sum rules,[31] regardless of $t$ and $\theta$. Similarly, normalized O $K$-edge XMCD spectra are identical to each other, meaning that the electronic structure contributing to the orbital magnetic moment of O, among the hybridized Ru 4$d$ $t_{2g}$-O 2$p$ states, does not change with $t$ and $\theta$. Figure 5(c) shows XMCD-$\mu_0 H$ curves measured at the Ru $M_3$-edge with $t = 2$ nm at 30 K for $\theta = 90°$ and $\theta = 20°$. There is no large anisotropy, but the magnetic moment at 1.92 T and the remanent magnetization at 0 T are slightly larger in the in-plane configuration, indicating weak in-plane magnetic anisotropy for the SRO film with $t = 2$ nm. To investigate more quantitatively the magnetic anisotropy in the SRO films with $t > 2$ nm, we plotted the $M_\parallel^{Ru}/M_\perp^{Ru}$ and $m_{orb,\parallel}^{O}/m_{orb,\perp}^{O}$ vs. $t$ at 30 K with a magnetic field $\mu_0 H$ of 1.92 T [Fig. 5(d)]. In Fig. 5(d), the data reported in Ref. 49 are also plotted for $t = 60$ nm. Here, $M_\perp^{Ru}$ and $m_{orb,\perp}^{O}$ are the total magnetic moment of Ru ($M^{Ru} = m_{orb}^{Ru} + m_{spin}^{Ru}$) at $\theta = 90°$ and the $m_{orb}^{O}$ at $\theta = 90°$, respectively. The $M_\parallel^{Ru}/M_\perp^{Ru}$ and $m_{orb,\parallel}^{O}/m_{orb,\perp}^{O}$ represent the magnetic anisotropy of Ru and O, respectively. A ratio larger than 1 indicates in-plane magnetic anisotropy. As reported in Ref. 49, at $t = 60$ nm, both $M_\parallel^{Ru}/M_\perp^{Ru}$ and $m_{orb,\parallel}^{O}/m_{orb,\perp}^{O}$ are 1.4, indicating the in-plane magnetic anisotropy with magnetic coupling between the Ru and O ions. In contrast, at $t = 2$ and 10 nm, both $M_\parallel^{Ru}/M_\perp^{Ru}$ and $m_{orb,\parallel}^{O}/m_{orb,\perp}^{O}$ are about 1.0-1.1 (almost isotropic), suggesting that the magnetic anisotropy is weakened by the presence of interface defects while Ru and O are still magnetically coupled. As reported in Ref. 50, in compressive-strained SRO/STO films, strong perpendicular magnetic anisotropy remained at $t = 10$ nm, where the magnetic moment in the perpendicular direction was about 1.8 times larger than in-plane. This difference indicates that the ultrathin tensile-strained SRO is more suitable for use as a soft magnet.

## IV. CONCLUSION

Our systematic investigation using magnetotransport, XAS, and XMCD measurements has shed light on the interplay between thickness, strain, and electronic structure in tensile-strained SRO films. Decreasing film thickness $t$ reveals a progressive localization of the Ru 4$d$ $t_{2g}$-O 2$p$ hybridized states, evidenced by the declining intensity of the coherent XAS peak. This localization, likely exacerbated by interfacial disorder, weakens the charge transfer-induced O 2$p$ orbital moment and the Ru 4$d$ magnetic moment, leading to a complete suppression of ferromagnetism at $t = 1$ nm. Furthermore, the increasing localization coincides with the emergence of insulating behavior for $t \leq 2$ nm. Strikingly, even below $t = 10$ nm, the in-plane magnetic anisotropy exhibits a pronounced reduction compared to thicker films, highlighting its sensitivity to the weakening of orbital hybridization under tensile strain. These findings offer valuable insights for engineering SRO-based electrodes in future electrical and spintronic devices. By carefully considering the delicate interplay between strain, thickness, and electronic structure, we can optimize SRO's behavior to achieve desired electrical and magnetic properties at the heterointerface, paving the way for innovative device architectures.

## ACKNOWLEDGMENTS
This work was partially supported by the Spintronics Research Network of Japan (Spin-RNJ). The experiment at SPring-8 was approved by the Japan Synchrotron Radiation Research Institute (JASRI) Proposal Review Committee (Proposal No. 2023B1441).

## CONFLICT OF INTEREST
The authors have no conflicts to disclose.

## AUTHORS' CONTRIBUTIONS
**Y. K. Wakabayashi**: Conceptualization (lead); Validation (lead); Investigation (lead); Supervision (lead); Writing – Original Draft (lead); Writing – Review & editing (lead). **M. Kobayashi**: Investigation (supporting); Supervision (supporting); Writing – Review & editing (supporting). **Y. Seki, K. Yamagami, T. Takeda, T. Ohkochi**: Investigation (supporting); Writing – Review & editing (supporting). **Y. Taniyasu**: Writing – Review & editing (supporting). **Y. Krockenberger**: Investigation (supporting); Writing – Review & editing (supporting). **H. Yamamoto**: Writing – Review & editing (supporting).

## DATA AVAIRABILITY
The data that support the findings of this study are available from the corresponding author upon reasonable request.

**REFERENCES**

[1] J. J. Randall, R. Ward, J. Am. Chem. Soc. **81**, 2629 (1959).

[2] L. Klein, J. S. Dodge, C. H. Ahn, J. W. Reiner, L. Mieville, T. H. Geballe, M. R. Beasley, A. Kapitulnik, J. Phys. Condens. Matter **8**, 10111 (1996).

[3] C. B. Eom, R. J. Cava, R. M. Fleming, J. M. Phillips, R. B. van Dover, J. H. Marshall, J. W. P. Hsu, J. J. Krajewski, W. F. Peck, Jr., Science **258**, 1766 (1992).

[4] G. Koster, L. Klein, W. Siemons, G. Rijnders, J. S. Dodge, C.-B. Eom, D. H. A. Blank, M. R. Beasley, Rev. Mod. Phys. **84**, 253 (2012).

[5] Y. K. Wakabayashi, Y. Krockenberger, T. Otsuka, H. Sawada, Y. Taniyasu, H. Yamamoto, Jpn. J. Appl. Phys. **62**, SA0801 (2023).

[6] D. Kan, M. Mizumaki, T. Nishimura, Y. Shimakawa, Phys. Rev. B **94**, 214420 (2016).

[7] D. E. Shai, C. Adamo, D. W. Shen, C. M. Brooks, J. W. Harter, E. J. Monkman, B. Burganov, D. G. Schlom, K.M. Shen, Phys. Rev. Lett. **110**, 087004 (2013).

[8] A. P. Mackenzie, J. W. Reiner, A. W. Tyler, L. M. Galvin, S. R. Julian, M. R. Beasley, T. H. Geballe, A. Kapitulnik, Phys. Rev. B **58**, R13318 (1998).

[9] Z. Li, S. Shen, Z. Tian, K. Hwangbo, M. Wang, Y. Wang, F. M. Bartram, L. He, Y. Lyu, Y. Dong, G. Wan, H. Li, N. Lu, J. Zang, H. Zhou, E. Arenholz, Q. He, L. Yang, W. Luo, P. Yu, Nat. Commun. **11**, 184 (2020).

[10] Y. Chen, D. L. Bergman, A. A. Burkov, Phys. Rev. B **88**, 125110 (2013).

[11] K. Takiguchi, Y. K. Wakabayashi, H. Irie, Y. Krockenberger, T. Otsuka, H. Sawada, S. A. Nikolaev, H. Das, M. Tanaka, Y. Taniyasu, H. Yamamoto, Nat. Commun. **11**, 4969 (2020).

[12] S. K. Takada, Y. K. Wakabayashi, Y. Krockenberger, T. Nomura, Y. Kohama, H. Irie, K. Takiguchi, S. Ohya, M. Tanaka, Y. Taniyasu, H. Yamamoto, npj Quantum Mater. **7**, 102 (2022).

[13] U. Kar, A. K. Singh, Y.-T. Hsu, C.-Y. Lin, B. Das, C.-T. Cheng, M. Berben, S. Yang, C.-Y. Lin, C.-H. Hsu, S. Wiedmann, W.-C. Lee, and W.-L. Lee, npj Quantum Mater. **8**, 8 (2023).

[14] W. Lin, L. Liu, Q. Liu, L. Li, X. Shu, C. Li, Q. Xie, P. Jiang, X. Zheng, R. Guo, Z. Lim, S. Zeng, G. Zhou, H. Wang, J. Zhou, P. Yang, Ariando, S. J. Pennycook, X. Xu, Z. Zhong, Z. Wang, J. Chen, Adv. Mater. **33**, 2101316 (2021).

[15] H. Boschker, T. Harada, T. Asaba, R. Ashoori, A. V. Boris, H. Hilgenkamp, C. R. Hughes, M. E. Holtz, L. Li, D. A. Muller, H. Nair, P. Reith, X. R. Wang, D. G. Schlom, A. Soukiassian, J. Mannhart, Phys. Rev. X **9**, 011027 (2019).

[16] K. S. Takahashi, A. Sawa, Y. Ishii, H. Akoh, M. Kawasaki, Y. Tokura, Phys. Rev. B **67**, 094413 (2003).

[17] L. Liu, Q. Qin, W. Lin, C. Li, Q. Xie, S. He, X. Shu, C. Zhou, Z. Lim, J. Yu, W. Lu, M. Li, X. Yan, S. J. Pennycook, J. Chen, Nat. Nanotechnol. **14**, 939 (2019).

[18] D. Toyota, I. Ohkubo, H. Kumigashira, M. Oshima, T. Ohnishi, M. Lippmaa, M. Takizawa, A. Fujimori, K. Ono, M. Kawasaki, H. Koinuma, Appl. Phys. Lett. **87**, 162508 (2005).

[19] Y. J. Chang, C. H. Kim, S.-H. Phark, Y. S. Kim, J. Yu, T. W. Noh, Phys. Rev. Lett. **103**, 057201 (2009).

[20] J. Xia, W. Siemons, G. Koster, M. R. Beasley, A. Kapitulnik, Phys. Rev. B **79**, 140407(R) (2009).

[21] X. Shen, X. Qiu, D. Su, S. Zhou, A. Li, D. Wu, J. Appl. Phys. **117**, 015307 (2015).

[22] S. K. Takada, Y. K. Wakabayashi, Y. Krockenberger, S. Ohya, M. Tanaka, Y. Taniyasu, H. Yamamoto, Appl. Phys. Lett. **118**, 092408 (2021).

[23] Q. Gan, R. A. Rao, C. B. Eom, J. L. Garrett, M. Lee, Appl. Phys. Lett. **72**, 978 (1998).

[24]C. U. Jung, H. Yamada, M. Kawasaki, Y. Tokura, Appl. Phys. Lett. **84**, 2590 (2004).
[25]A. T. Zayak, X. Huang, J. B. Neaton, K. M. Rabe, Phys. Rev. B **74**, 094104 (2006).
[26]D. Kan, R. Aso, H. Kurata, Y. Shimakawa, J. Appl. Phys. **113**, 173912 (2013).
[27]Y. K. Wakabayashi, S. K. Takada, Y. Krockenberger, Y. Taniyasu, H. Yamamoto, ACS Applied Electronic Materials **3**, 2712 (2021).
[28]Y. K. Wakabayashi, T. Otsuka, Y. Krockenberger, H. Sawada, Y. Taniyasu, H. Yamamoto, APL Mater. **7**, 101114 (2019).
[29]Y. K. Wakabayashi, T. Otsuka, Y. Krockenberger, H. Sawada, Y. Taniyasu, H. Yamamoto, npj Comput. Mater. **8**, 180 (2022).
[30]Y. K. Wakabayashi, T. Otsuka, Y. Krockenberger, H. Sawada, Y. Taniyasu, H. Yamamoto, APL Machine Learning **1**, 026104 (2023).
[31]B. T. Thole, P. Carra, F. Sette, G. van der Laan, Phys. Rev. Lett. **68**, 1943 (1992).
[32]C. T. Chen, Y. U. Idzerda, H. -J. Lin, N. V. Smith, G. Meigs, E. Chaban, G. H. Ho, E. Pellegrin, F. Sette, Phys. Rev. Lett. **75**, 152 (1995).
[33]J. Stohr, H. Konig, Phys. Rev. Lett. **75**, 3748 (1995).
[34]Y. K. Wakabayashi, Y. Nonaka, Y. Takeda, S. Sakamoto, K. Ikeda, Z. Chi, G. Shibata, A. Tanaka, Y. Saitoh, H. Yamagami, M. Tanaka, A. Fujimori, R. Nakane, Phys. Rev. B **96**, 104410 (2017).
[35]Y. K. Wakabayashi, S. K. Takada, Y. Krockenberger, K. Takiguchi, S. Ohya, M. Tanaka, Y. Taniyasu, H. Yamamoto, AIP advances **11**, 035226 (2021).
[36]S. K. Takada, Y. K Wakabayashi, Y. Krockenberger, H. Irie, S. Ohya, M. Tanaka, Y. Taniyasu, H. Yamamoto, Phys. Rev. Mater. **7**, 054406 (2023).
[37]J. Snoek, H. Larochelle, R. P. Adams, paper presented at Advances in Neural Information Processing Systems **2012**, http://papers.nips.cc/paper/ 4522-practical-bayesian-optimization.
[38]Y. K. Wakabayashi, Y. Krockenberger, N. Tsujimoto, T. Boykin, S. Tsuneyuki, Y. Taniyasu, H. Yamamoto, Nat. Commun. **10**, 535 (2019).
[39]Y. Senba, H. Ohashi, Y. Kotani, T. Nakamura, T. Muro, T. Ohkochi, N. Tsuji, H. Kishimoto, T. Miura, M. Tanaka, M. Higashiyama, S. Takahashi, Y. Ishizawa, T. Matsushita, Y. Furukawa, T. Ohata, N. Nariyama, K. Takeshita, T. Kinoshita, A. Fujiwara, M. Takata, S. Goto, AIP Conf. Proc. **1741**, 030044 (2016).
[40]T. Nakamura, T. Muro, F. Z. Guo, T. Matsushita, T. Wakita, T. Hirono, Y. Takeuchi, K. Kobayashi, J. Electron Spectrosc. Relat. Phenom. **144-147**, 1035 (2005).
[41]A. Rastogi, M. Brahlek, J. M. Ok, Z. Liao, C. Sohn, S. Feldman, H. N. Lee, APL Mater. **7**, 091106 (2019).
[42]D. B. Kacedon, R. A. Rao, C. B. Eom, Appl. Phys. Lett. **71**, 1724 (1997).
[43]M. Tsunoda, Y. Komasaki, S. Kokado, S. Isogami, C. C. Chen, M. Takahashi, Appl. Phys. Express **2**, 083001 (2009).
[44]J. Okamoto, T. Okane, Y. Saitoh, K. Terai, S.-I. Fujimori, Y. Muramatsu, K. Yoshii,1 K. Mamiya, T. Koide, A. Fujimori, Z. Fang, Y. Takeda, M. Takano, Phys. Rev. B **76**, 184441 (2007).
[45]K. Terai, K. Yoshii, Y. Takeda, S. I. Fujimori, Y. Saitoh, K. Ohwada, T. Inami, T. Okane, M. Arita, K. Shimada, H. Namatame, M. Taniguchi, K. Kobayashi, M. Kobayashi, A. Fujimori, Phys. Rev. B **77**, 115128 (2008).
[46]Y. K. Wakabayashi, M. Kobayashi, Y. Takeda, K. Takiguchi, H. Irie, S.-i. Fujimori, T. Takeda,

R. Okano, Y. Krockenberger, Y. Taniyasu, H. Yamamoto, Phys. Rev. Mater. **5**, 124403 (2021).

[47]M. Takizawa, D. Toyota, H. Wadati, A. Chikamatsu, H. Kumigashira, A. Fujimori, M. Oshima, Z. Fang, M. Lippmaa, M. Kawasaki, H. Koinuma, Phys. Rev. B **72**, 060404(R) (2005).

[48]H. Jeong, S. G. Jeong, A. Y. Mohamed, M. Lee, W. Noh, Y. Kim, J. S. Bae, W. S. Choi, D. Y. Cho, Appl. Phys. Lett. **115**, 092906 (2019).

[49] Y. K. Wakabayashi, M. Kobayashi, Y. Takeda, M. Kitamura, T. Takeda, R. Okano, Y. Krockenberger, Y. Taniyasu, H. Yamamoto, Phys. Rev. Mater. **6**, 094402 (2022).

[50]Y. K. Wakabayashi, M. Kobayashi, Y. Seki, Y. Kotani, T. Ohkochi, K. Yamagami, M. Kitamura, Y. Taniyasu, Y. Krockenberger, H. Yamamoto, arXiv 2309.05228 (2023).

[51]J. Igarashi, K. Hirai, Phys. Rev. B **53**, 6442 (1996).

**Figures and figure captions**

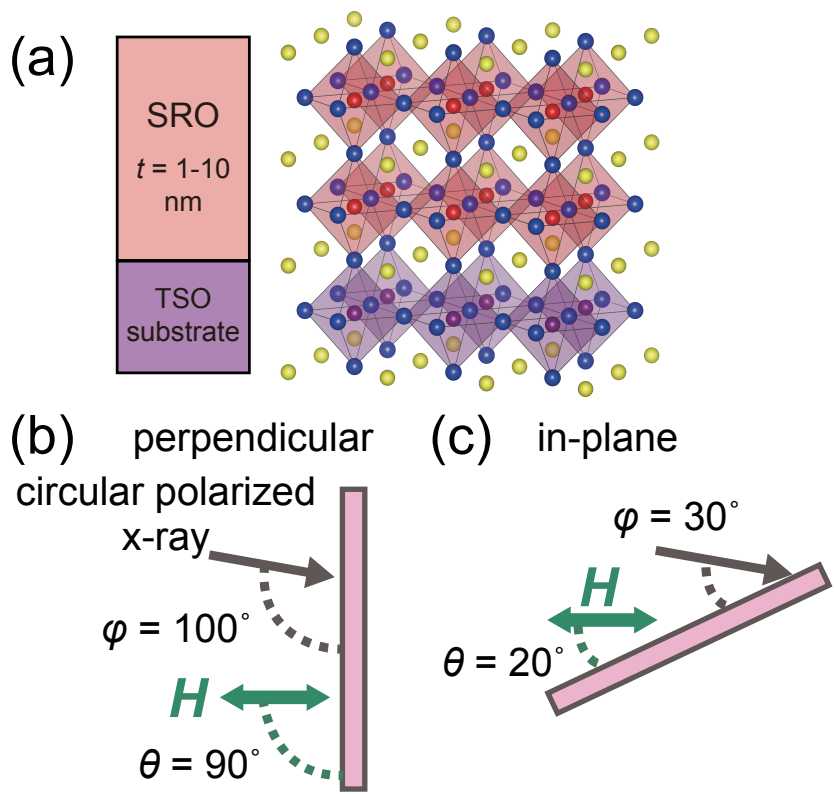

**Fig. 1.** Schematic diagrams depict the sample and crystal structures of the SRO films. In the crystal image, strontium, oxygen, ruthenium, and titanium are represented by yellow, blue, red, and purple spheres, respectively. Illustrations outline the measurement configurations for (b) perpendicular and (c) in-plane XAS and XMCD measurements.

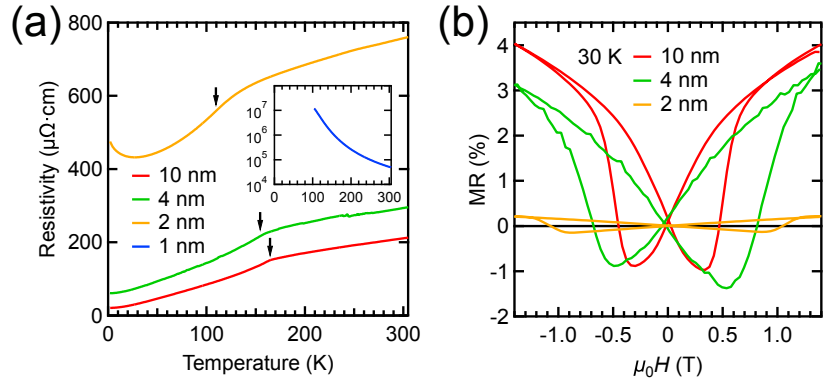

**Fig. 2.** (a) Temperature dependence of resistivity $\rho$ on thickness $t$ (=1–10 nm). Arrows indicate kinks. The inset in (a) shows data plotted with log scale for $t = 1$ nm. (b) Thickness $t$ dependence of MR $(\rho(B)-\rho(0\,T))/\rho(0\,T)$ at 30 K with $\mu_0 H$ applied in the out-of-plane [110] direction of the TSO substrate.

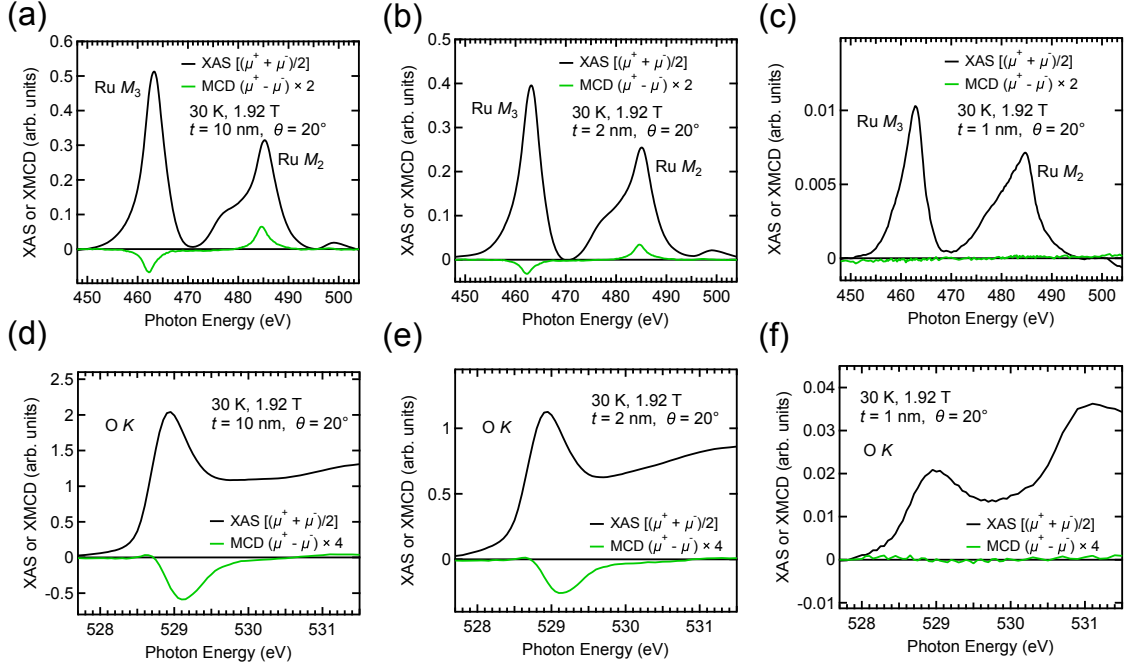

**Fig. 3.** (a)–(c) XAS and XMCD spectra at the Ru $M_{2,3}$ edge with $t =$ (a) 10, (b) 2, and (c) 1 nm at 30 K with a magnetic field $\mu_0 H$ of 1.92 T at $\theta = 20°$. (d)–(f) XAS and XMCD spectra at the O $K$ edge with $t =$ (d) 10, (e) 2, and (f) 1 nm.

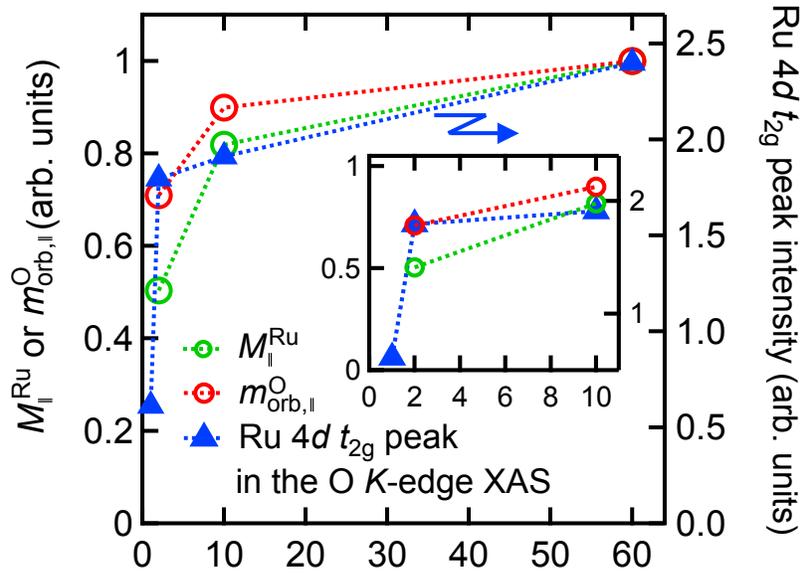

**Fig. 4.** Thickness dependence of the Ru $4d$ $t_{2g}$ peak in the O $K$-edge XAS spectrum normalized at 531 eV, $M_\parallel^{Ru}$, and $m_{orb,\parallel}^O$ at 30 K under a magnetic field $\mu_0 H$ of 1.92 T. The data reported in Ref. 49 are plotted for $t = 60$ nm.

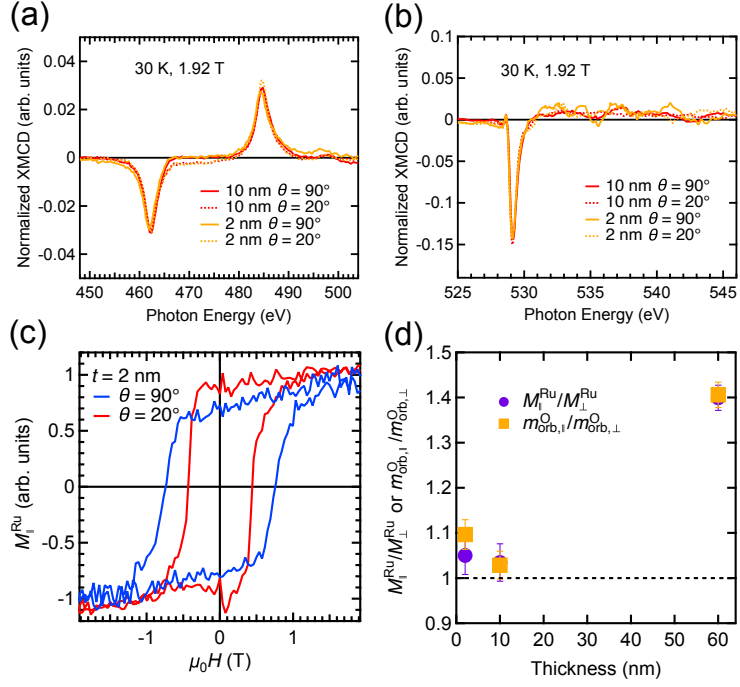

**Fig. 5.** (a),(b) Ru $M_{2,3}$-edge XMCD spectra normalized at 462.2 eV (a) and O $K$-edge XMCD spectra normalized at 529.1 eV (b) with $t$ = 10 and 2 nm at $\theta = 90°$ and $\theta = 20°$. (c) XMCD-$\mu_0 H$ curves measured at the Ru $M_3$-edge with $t$ = 2 nm at 30 K and $\theta = 90°$ and $\theta = 20°$. (d) Thickness dependence of the $M_\parallel^{Ru}/M_\perp^{Ru}$ and $m_{orb,\parallel}^O/m_{orb,\perp}^O$ at 30 K under a magnetic field $\mu_0 H$ of 1.92 T. In (d), the data reported in Ref. 49 are also plotted for $t$ = 60 nm.